\documentclass[12pt]{article}
\usepackage{xspace,amssymb,amsmath,helvet,graphicx}
\begin{document}
% useful commands
\newcommand{\dee}{\,\mbox{d}}
\newcommand{\naive}{na\"{\i}ve }
\newcommand{\Naive}{Na\"{\i}ve }
\newcommand{\eg}{e.g.\xspace}
\newcommand{\ie}{i.e.\xspace}
\newcommand{\pdf}{pdf.\xspace}
\newcommand{\etc}{etc.\@\xspace}
\newcommand{\PhD}{Ph.D.\xspace}
\newcommand{\MSc}{M.Sc.\xspace}
\newcommand{\BA}{B.A.\xspace}
\newcommand{\MA}{M.A.\xspace}
\newcommand{\role}{r\^{o}le}
\newcommand{\signoff}{\hspace*{\fill} Rose Baker \today}
% entry environment
\newenvironment{entry}[1]%
{\begin{list}{}{\renewcommand{\makelabel}[1]{\textsf{##1:}\hfil}%
\settowidth{\labelwidth}{\textsf{#1:}}%
\setlength{\leftmargin}{\labelwidth}
\addtolength{\leftmargin}{\labelsep}
\setlength{\itemindent}{0pt}
}}%
{\end{list}}
\title{A new measure of treatment effect for random-effects meta-analysis of comparative binary outcome data}
\author{Rose Baker\\School of Business\\University of Salford, UK\\Dan Jackson\\Statistical Innovation Group\\Advanced
Analytics Centre\\AstraZeneca, Cambridge, UK\footnote{This work was done while Dan was working at the MRC/BSU, Cambridge}}
\maketitle
\begin{abstract}
Comparative binary outcome data are of fundamental interest in  statistics and are often pooled in meta-analyses.
Here we examine the simplest case where  for each study there are two patient groups and a binary event of interest, giving rise to a series of $2 \times 2$ tables.
A variety of measures of treatment effect are then available and are conventionally used in meta-analyses, such as the odds ratio, the risk ratio and the risk difference.
Here we propose a new type of measure of treatment effect for this type of data that is very easily interpretable by lay audiences.
We give the rationale for the new measure and we present three contrasting methods for computing its within-study variance so that it can be used in conventional meta-analyses. We then develop three alternative methods for random-effects meta-analysis that use our measure and we apply our methodolgy to some real examples. We conclude that our new measure is a fully viable alternative to existing measures. It has the advantage that its interpretation is especially simple and direct, so that its meaning can be more readily understood by those with little or no formal statistical training. This may be especially valuable when presenting `plain language summaries', such as those used by Cochrane.
\end{abstract}


\section{Introduction}
In medicine, we often want to measure the effect of a treatment, usually a drug or a medical intervention,
and in epidemiology, we often wish to measure the effect of exposure to some health hazard. Multiple studies that provide relevant data are often available which may then be pooled in meta-analyses. We consider the case of comparative binary outcome data, where for each study there are two patient groups and a binary event of interest, such as death. Interest then lies in determining which patient group is more likely to experience the event.  Most studies of this type are parallel studies, with a control (or placebo) group and a treatment group. Occasionally, studies are paired, where each patient acts as their own control.  Here we focus on parallel studies and the estimation of an appropriate treatment effect for the resulting series of $2 \times 2$ tables. We will use random-effects meta-analyses for this purpose, so that the possibility of between-study heterogeneity is included in our modelling. If the strong assumption of homogeneity is made then common-effect models may be used instead, which greatly simplifies the statistical methods required.

Although the $2 \times 2$ table is a particularly simple and common data structure, the issues relating to the analysis of this type of data are subtle. In particular, there are several issues that should be considered when determining an appropriate measure of treatment effect for this type of data. For example, there is an important distinction between relative and absolute measures (Deeks, 2002). The Cochrane handbook (Higgins and Green, 2011), its section 9.7, 
under the heading of sensitivity analyses, asks `for dichotomous outcomes, should odds ratios, risk ratios or risk differences be used?'. We therefore have three conventional measures of treatment effect for performing meta-analyses involving comparative binary data but all of these measures can be difficult for lay audiences to interpret. For example, according to Davies {\em et al} (1998), `odds ratios are hard to comprehend directly', and  Grimes and Schulz (2008) state that `for most clinicians, odds ratios will remain ... well, odd'.  Risk ratios and differences are probably easier to interpret but Schechtman (2002) explains that these measures also have their disadvantages, where the problems stem from the fact that the same risk difference or risk ratio might have very different implications depending on the baseline risk. In order to make the risk difference more interpretable its reciprocal, the number needed to treat (Nuovo, Melnikow and Chang, 2002), has been proposed; if the estimated risk difference indicates that the treatment is not beneficial relative to the control then this measure is interpreted as the number needed to harm. The number needed to treat is a very appealing and intuitive measure for non-statisticians to interpret but has serious statistical difficulties (Hutton, 2002 and 2010). The poor statistical properties of the number needed to treat are a consequence of the fact that it is undefined under the null hypothesis where the probability of an event is the same in both groups. 

Our aim here is to develop a new measure of treatment effect for comparative binary outcome data that, like the risk difference, takes values in the interval [-1, 1], and is easily interpretable by non-statisticians. The proposed measure will have a very simple and intuitively appealing interpretation, along the lines of the number needed to treat.
The effect will be zero under the null.
Minus one  will indicate that no patients in the treatment group experience the event (but some in the control do), and plus one will indicate that all patients in the  treatment group  experience the event (but some in the control group do not). The probability that patients in the treatment group experience the event will be  a monotonically increasing function of the treatment effect, for a given probability in the control group.
A  related idea to ours is the proposal of Mirzazadah, Malekinejad and Kahn (2015), the `relative risk reduction of an undesirable outcome'. This is a simple transformation of the relative risk, which our measure generalises. The disadvantages of our measure are that it will necessarily be unfamiliar, and so appear strange to statisticians, and that it is not differentiable (but is continuous) at the null. However this is not a serious statistical difficulty in practice. Our hope is that our ideas could be used to make meta-analyses and systematic reviews, and indeed statistical analyses more generally, more accessible to those with little or no formal statistical training. We return to this issue in the discussion.

The rest of the paper is set out as follows. In section 2 we summarise existing measures, develop our new measure, describe three methods to compute its variance and develop an accurate approximation to its distribution.  In section 3 we develop three random-effects models for meta-analysis that use our new measure. The first of these models simply uses the conventional random-effects model to describe the outcome data but the second two models are novel and are motivated by the desire to make more accurate inferences using our new measure. In section 4 we apply our new methods to three real meta-analyses. We conclude in section 5 with a short discussion.

\section{A new measure of treatment effect and its properties}
In this section we summarise the most popular existing measures of treatment effect for analyzing $2 \times 2$ tables, develop our new measure and describe its properties. 
\subsection{Existing measures}
In this section we describe methods for a single $2 \times 2$ table and in section three we will develop methods for the random-effects meta-analysis of multiple tables. 
For comparative binary data, a variety of measures of treatment effect $\theta$ are currently available.
We will use $\theta$ to denote the treatment effect, where the type of treatment effect that this refers to will be obvious from the context, and ultimately we will use $\theta$ to denote our new measure.

Let $p$ denote the probability of an event in the control group and let $q$ denote the probability of an event in the treatment group. All the measures that follow are suitable functions of $p$ and $q$. We will see below that our new measure is another such function, but one where a simple causal explanation can be used to communicate its meaning.  

In the context of meta-analysis, Hartung, Knapp and Sinha (2008) also give an account of many of the established measures that follow, to which the reader is referred for more details.

\subsubsection{The odds ratio.}

A very popular measure of the relative treatment effect is the odds ratio, $\frac{q/(1-q)}{p/(1-p)}$. This, and all the quantities that follow, are estimated by replacing $p$ and $q$ with their estimates, $\hat{p}$ and $\hat{q}$, which are the observed proportions of patients that experience the event of interest in each group. Analyses are usually performed on the log scale so that the  log odds-ratio, $\ln\{\frac{q/(1-q)}{p/(1-p)}\}$ is used in analysis. Inferences may then be back-transformed to the odds scale. The log-odds ratio is undefined when $p=0,1$ or $q=0,1$ and halves or some other quantity are usually added to all entries of the $2 \times 2$ table prior to analysis to avoid this problem when there are zeros. This also applies to the relative risk when either probability is zero. The odds ratio may take any non-negative value, and the log odds ratio may take any value $(-\infty,\infty)$. As explained in the introduction, the odds ratio is not an easily interpretable quantity for many consumers of statistical analyses, and its use is usually motivated by its good statistical properties and connections with other standard statistical methods, such as logistic regression.

The odds ratio is label-invariant and the log-odds ratio simply changes sign when `good' and `bad' outcomes are switched; we use the shorthand $p\rightarrow 1-p, q \rightarrow 1-q$ to indicate this change. Similarly the log-odds ratio switches sign when treatment and control groups are interchanged, for which we use $p \leftrightarrow q$. The invariance property of the log-odds ratio is another reason why it is often preferred over some of the measures that follow.

\subsubsection{The relative risk.}
The relative risk of an event $q/p$, and  the relative risk of not experiencing the effect $(1-q)/(1-p)$, are also commonly used relative measures of a treatment effect. These measures are more easily interpretable than the odds ratio. As with the odds ratio, analyses are usually performed on the log risk scale.
The relative risk of the event and not experiencing the event are not the same, so the relative risk is not label-invariant when $p \rightarrow 1-p$, $q \rightarrow 1-q$, but is
invariant under  $ p \leftrightarrow q$.

\subsubsection{The risk difference.}

The risk difference $q-p$ is an absolute measure of treatment effect, that can take values in the interval [-1,1]. It has the problem that if $q=p+\theta$, for some values of $p$, $q$ will lie outside $[0,1]$ for $\theta \in [-1,1]$. In order to make this measure more interpretable,
 its reciprocal $1/(q-p)$, the number needed to treat,  has been proposed. However as explained in the introduction, the number needed to treat has been criticised because of its poor statistical properties which stem from the fact that it is undefined under the null where $p=q$.

\subsubsection{The arcsine difference.}
The arcsine difference $\sin^{-1}\sqrt{q}-\sin^{-1}\sqrt{p}$ is a risk difference with the variances of $p$ and $q$ stabilised using a variance stabilising transformation. This measure has been proposed by R\"{u}cker {\em et al} (2009), particularly in situations where the event of interest is rare. The arcsine can take values in the interval $[-\pi/2,\pi/2]$.

\subsubsection{Families of measures of treatment effects.}
Jackson, Baker and Bowden (2013) propose a family of treatment effects of the form $T(q)-T(p)$ that includes many of the measures described above and can be used in a sensitivity analysis. The transformation $T$ used by Jackson {\em et al}. was inspired by the one proposed by Aranda-Ordaz (1981) that can also be used for this purpose.

\subsection{New measure: The `GRRR'}
One observation from section 2.1 is that a wide variety of measures of treatment effect have been proposed and used in analysis. In particular, some of these are routinely used in meta-analyses.  Other than the number needed to treat, which has poor statistical properties, all measures are, for one reason or another, hard for lay audiences to interpret. In this section we develop another measure, the `Generalised Relative Risk Reduction', which gives rise to the whimsical acronym `GRRR'.
Our new measure has a very simple interpretation, as is also the case for the number needed to treat, but the GRRR has  more acceptable statistical properties.
The key concept is that as $\theta$ increases from $-1$ one towards zero, an increasing (from zero) proportion of those who experience the event under the placebo would also experience this under treatment until $\theta=0$ when $q=p$. Also as $\theta$ subsequently increases towards unity, an increasing proportion of those not experiencing the event under the placebo would experience this under the treatment. When $\theta=1$,  all patients in the treatment group experience the event.

We begin by considering the probability $q$ as a function of $p$ and the new measure $\theta$. We define $q=q(p, \theta)$ so as to ensure that $\theta$ represents a meaningful and easily interpretable quantity.
We will define our measure  differently for $\theta < 0$ and $\theta > 0$ whilst ensuring that $\theta$ is easily interpreted in either case. We then put these definitions together to define our measure.
We require that $\theta=0$ is equivalent to $q=p$, and also that $\theta \in [-1,1]$ acts continuously on  $q$, so that $\partial q/\partial \theta > 0$. Finally, we will ensure that $\theta= \pm 1$ represents the greatest possible treatment effects.

\subsubsection{The case where $q > p$.}
 If $q > p$, so that the event is at least as likely in the treatment group as in the control (and is not certain in the control group) we define $q=p+\theta(1-p)$, where $0 < \theta<1$. This can be interpreted as meaning that, in addition to the proportion who experienced the event in the control group (and who would also have experienced the event if they were instead in the treatment group),
a further proportion $\theta$ of those who would not experience the event in the control group would have experienced this event if they were in the treatment group. For example, $\theta=0.6$ can be interpreted as meaning that 60\% of patients who do not experience the event in the control group would have experienced the event if they were in the treatment group.
This may be more easily understood as $1-q=(1-\theta)(1-p)$, so that the proportion $1-p$ not experiencing the event shrinks by a factor of $1-\theta$. When $\theta=1$ we have that $q=1>p$, so that the event is certain in the treatment group whilst not certain in the control group.

\subsubsection{The case where $q < p$.}
If  $q < p$, so that the event is less likely in the treatment group (and is not certain in the treatment group) we define $q=(1+\theta)p$ where $-1<\theta<0$. This can be interpreted as meaning that, if $\theta$ is negative, a proportion $1+\theta$ of those who would experience the event in the control group would also experience the event if they were in the treatment group (and all those who would not experience the event in the control group would also not experience the event if they were in the treatment group). For example, $\theta=-0.6$ can be interpreted as meaning that  40\% of patients who experience the event in the control group would also experience this event if they were in the treatment group. When $q=0$ and $p>0$ we have $\theta=-1$, so that $\theta=-1$ indicates that the event is impossible in the treatment group whilst  being possible in the control group.
\subsubsection{The case where $q = p$.}
If $p=q$ then $\theta=0$ and we could take $q=p+\theta(1-p)$, as in the case where $q>p$, or $q=p(1+\theta)$, as in the case where $q<p$. We take the former option but this makes no material difference. Hence we arbitrarily use the definition for the case where  $q > p$ to apply slightly more generally to $q \ge p$.

\subsubsection{Putting these three cases together and defining our new measure.}
The direct and easily interpretable nature of our new measure $\theta$ is now apparent, because it may be interpreted as simply modifying the response of a subset of the control group patients in order to produce the treatment group probabilities. We have used the causal language that the `event would have been  different'  for some easily identified proportions of patients in the control group `if they had instead been in the treatment group' when motivating our measure. However we will see below that the measure $\theta$ is just another suitable function of $p$ and $q$ that can be used to measure the treatment effect. Other explanations of why $p$ and $q$ take their values, and so result in a particular value of $\theta$, are of course also possible and more likely than the simple minded causal explanations that we have used to motivate the measure. However our intention is to use this causal explanation for lay audiences to explain one reason (of many) for the treatment effect observed. We are also able to communicate the uncertainty in our estimates using this language, as we demonstrate for some of our meta-analyses  below.

Putting the three cases together, our proposed new treatment effect is defined by the function
\begin{equation}
q=\left\{\begin{array}{ll}
(1+\theta)p & \text{if $q < p$}\\
p+\theta (1-p)& \text{if $q \ge p$}.
\end{array}
\right. \label{eq:qdef}\end{equation}

Succinctly, the probability of an event in the treatment group is
\begin{equation}q=(1+\min(\theta,0))p+\max(\theta,0)(1-p),\label{eq:theta}\end{equation}
or $q-p=\min(\theta,0)p+\max(\theta,0)(1-p)$.
Writing $\min(\theta,0)=(\theta-|\theta|)/2,\, \max(\theta,0)=(\theta+|\theta|)/2$, we have the alternative form
\[q-p=\theta/2-|\theta|(p-1/2).\]
 Solving (\ref{eq:qdef}) for $\theta$ gives rise to the definition of the GRRR of
\begin{equation}
\label{eq2}
\theta=\left\{\begin{array}{ll}
(q/p)-1 & \text{if ${q} < {p}$}\\
1-(1-q)/(1-p) & \text{if ${q} \ge {p}$}.
\end{array}
\right. \end{equation}
The nature of our generalised relative risk ratio is most evident from (\ref{eq2}): if ${q} < {p}$ then it is the relative risk minus one, and if ${q} \ge {p}$ then it is one minus the relative risk of not experiencing the event. Our measure is therefore a type of generalised relative risk ratio, hence its name. Although we motivated it without reference to risk ratios we can see now that  it can be expressed directly in terms of them. Therefore our measure is closely related to methods that all statisticians will be familiar with, and is not such a radical departure from conventional methods as it may at first appear. More succinctly we can write
\begin{equation}{\theta}=\frac{{q}-{p}}{{p}+(1-2{p})H({q}-{p})},\label{eq:true}\end{equation}
where  $H$ is the Heaviside step function, such that $H(x)=1$ if $x > 0$, else zero if $x < 0$, and $H(0)=1/2$; the case $p=q$ has been correctly specified as $\theta=0$ for all $p,q \in [0,1]$. Equation (\ref{eq:true}) appears to be an unusual candidate for a measure of treatment effect but it is  a convenient way performing the calculation. We  substitute the estimates $\hat{p}, \hat{q}$, which are just the observed proportions of events in each treatment group, into (\ref{eq:true}) to produce the estimated effect $\hat{\theta}$. Hence we can  write
\begin{equation}
\label{eq2a}
\hat{\theta}=\left\{\begin{array}{ll}
(\hat{q}/\hat{p})-1 & \text{if ${\hat{q}} < {\hat{p}}$}\\
1-(1-\hat{q})/(1-\hat{p}) & \text{if ${\hat{q}} \ge {\hat{p}}$}.
\end{array}
\right. \end{equation}
and
\begin{equation}{\hat{\theta}}=\frac{\hat{q}-\hat{p}}{\hat{p}+(1-2\hat{p})H(\hat{q}-\hat{p})},\label{eq:est}\end{equation}
Although $\theta$ was motivated using risk ratios, the numerators in (\ref{eq:true}) and (\ref{eq:est}) are the true and estimated risk differences, respectively. Hence these forms shows how the proposed measure relates to this other well known measure of treatment effect.

\begin{figure}
\centering
\makebox{\includegraphics[width=0.45\textwidth]{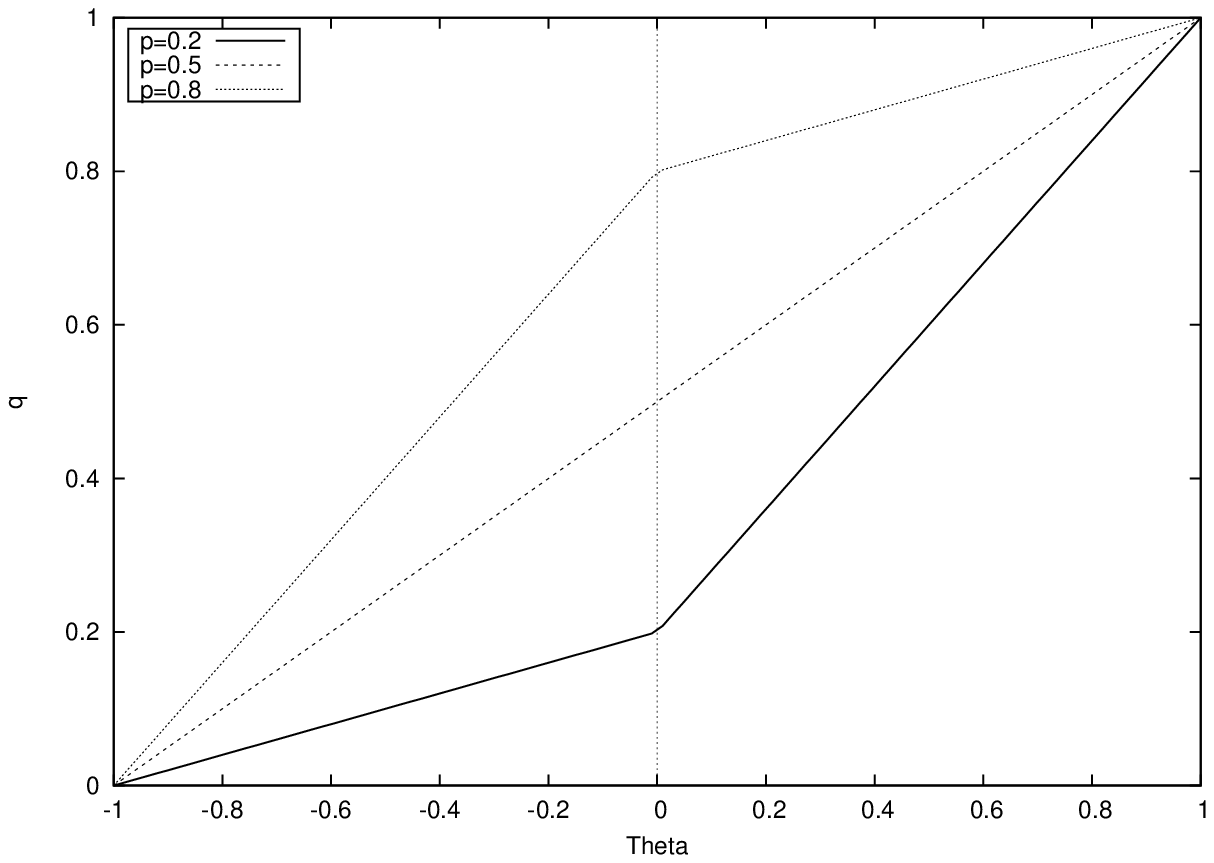}}
\makebox{\includegraphics[width=0.45\textwidth]{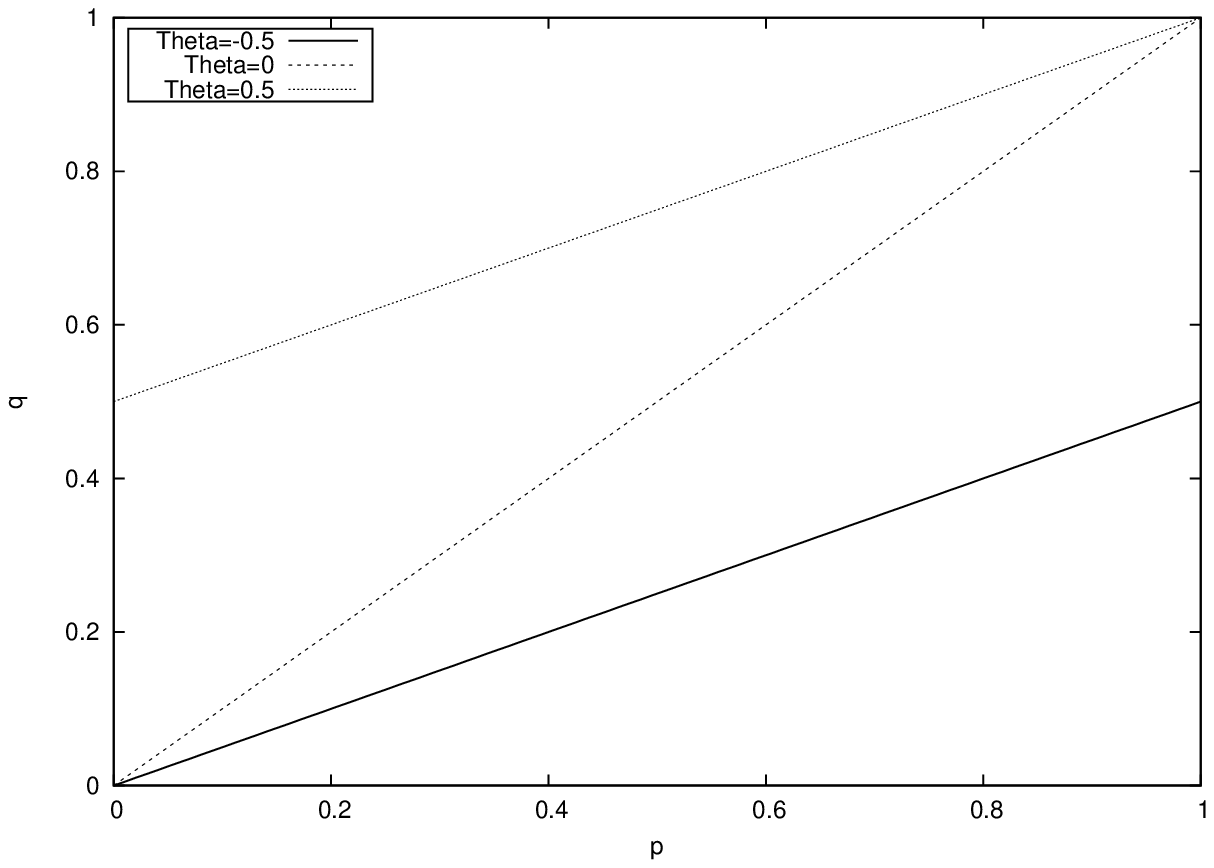}}
\caption{\label{figa} Left: Equation (\ref{eq:theta}) showing $q$ plotted against $\theta$ for several values of the control group probability $p$. Right: Equation (\ref{eq:theta}) showing $q$ plotted against $p$ for several values of the treatment effect $\theta$. }
\end{figure}

A physical way to illustrate our  measure, for those who think visually rather than numerically, could be to take a vessel such as a bottle with a long neck, the same length as the body, where in the neck the cross sectional area $p$ changes to $1-p$.
The total capacity of the bottle is a unit volume.
The body of the bottle is sunk into the ground, with the bottom at $-1$ and the top of the body at ground level, so that the top is at $+1$. Water is poured into the bottle, and the height of the water level is $\theta$.
The volume of water is $q$, so below ground level $q=(1+\theta)p$. At ground level, $q=p$ and $\theta=0$, and in the neck, $q=p+\theta(1-p)$.
The measure is shown in figure \ref{figa}.

\subsubsection{Properties of the proposed measure }
The GRRR is `label-invariant', under  $p\rightarrow 1-p, q \rightarrow 1-q$. This type of symmetry or label-invariance is  thought desirable, because
without it conclusions would depend on whether we looked at the proportion of patients recovering, or the proportion not recovering. However, the widely used relative risk does not possess this property. Deeks (2002) however notes that the `natural' choice out of `good' or `bad' relative risk usually fits the data better, so this lack of invariance is not a serious problem in practice.

Our measure is not label invariant with regard to switching the treatment and control
labels, when $p \leftrightarrow q$. This is a consequence of motivating the measure by a consideration of simple causal implications of what would have happened to patients in the control group if they had instead have been in the treatment group: if we instead apply these causal implications to what would have happened to patients in the treatment group it is immediately obvious that
different inferences for $\theta$ will be made. In situations where one group is the `placebo' or `unexposed group, as is the case for all the examples in this paper, then it is much more  natural to ask the question of what would have happened if patients in this group were treated or exposed, rather than ask this question `the other way round'.

However in situations where two treatments `A' and `B' are compared  there is no such natural treatment ordering. To help the lay audience understand the relative merits of two such treatments, it may therefore be useful to present two analyses, the first where `A' is the control and `B' is the treatment, and then vice versa. The results can then be phrased by explaining the implications of the treatments if some patients in group A were instead in group B, and then also if some patients were in group B instead of group A.
For showing that the treatments do not give identical results, a label-invariant significance test could be  performed.

The GRRR is not differentiable at $\theta=0$. This is a direct consequence of the fact that it is defined differently for positive and negative values. This is an undesirable property but we have not found it to be a serious issue in practice. However this would be a problem for any further methods that, for example, involved taking a Taylor series expansion that $\theta=0$. In any case, this is a much less serious issue than the one presented by the GRRR's main competitor for an easily interpreted measure of treatment effect, the number needed to treat, which is undefined when there is no treatment effect.

\subsection{Computing the variance  of the estimated treatment effect}
In order to make inferences using the proposed measure using a normal approximation (for example when using the conventional random-effects model for meta-analysis, see our method 1 below) we need methods for calculating the variance $\sigma^2$ of $\hat{\theta}$. We use the notation
in table \ref{tab:1}, so that for example  we can write $\hat{p}=n_{11}/N_1$, $\hat{q}=n_{12}/N_2$. We propose three different ways to compute $\sigma^2$.

\begin{table}[h]
\begin{tabular}{|l|l|l|} \hline
event$\downarrow$,group$\rightarrow$& Control&Treatment \\ \hline
Yes & $n_{11} $ & $n_{12} $ \\ \hline
No & $n_{21} $ & $n_{22} $ \\ \hline
Total&$N_1$ & $N_2$  \\ \hline
\end{tabular}
\caption{\label{tab:1}Notation for $2 \times 2$ tables; columns are the group, rows the event, \eg successful or unsuccessful.}
\end{table}

\subsubsection{Analytical calculation of the variance.}
For typical study sizes available,
complete enumeration is simplest and most accurate. R\"ucker  {\it et al} (2009) also consider this possibility for the arcsine difference, and refer to this as `analytical calculation of the variance' and we adopt their terminology here. Writing $P_i$ to denote the binomial probability of $i$ control group responses, and $Q_j$ to denote the probability of $j$ treatment group responses, with
$P_i={N_1 \choose i}\hat{p}^i(1-\hat{p})^{N_1-i}$,
$Q_j={N_2 \choose j}\hat{q}^i(1-\hat{q})^{N_2-j}$, we have
\begin{equation}\text{E}(\theta^m)=\sum_{i=0}^{N_1}\sum_{j=0}^{N_2} \theta_{ij}^m P_iQ_j,\label{eq:var}\end{equation}
where $\theta_{ij}$ is the value of $\theta$ corresponding to $\hat{\hat{p}}=i/N_1$, $\hat{\hat{q}}=j/N_2$. From these calculations $\sigma^2=\text{E}(\theta^2)-(\text{E}(\theta))^2$.

For most sample sizes encountered in practice, it is currently perfectly feasible to compute (\ref{eq:var}).  The most efficient and robust way is to compute probabilities recursively, starting at the mode and continuing both up and  down to very low probabilities. The correct scale factor for the probabilities is found by requiring that they sum to unity.
As is the case with the next two methods, this approach provides only an approximate variance because it `plugs in'  the point estimates $\hat{p}$ and $\hat{q}$ instead of using the true (unknown) values $p, q$.

\subsubsection{Monte Carlo calculation of the variance.}
An alternative method is to use parametric bootstrapping,
where we simulate many binomial realizations from $P_i$ and $Q_j$, calculate $\hat{\theta}$ and the
variance of these bootstrap replications gives the required variance.
This method is attractive in situations where $N_1$ and/or $N_2$ are large, so that the enumeration
required in the previous method is less feasible.

\subsubsection{An approximate formula for the variance.}
The appendix give an approximate formula for the variance $\sigma^2$. This is based on the usual Taylor series expansion and normal approximation for the logged relative risk. The resulting formula requires use of the standard normal cumulative distribution but is very fast to compute and has been found to work very well when $0.1 < p < 0.9$, $0.1 < q < 0.9$, and $N_1 > 100, N_2 > 100$. The analysis in the appendix treats the cases where $q>p$ and $q<p$ separately, so that the non-differentiability at $\theta=0$ does not result in any difficulties for the approximations used.

\subsection{A split lognormal approximation for the distribution of $\hat{\theta}$}
A normal approximation for $\hat{\theta}$ can be used directly for making inferences about $\theta$  and can be anticipated to be adequate in situations where such an approximation for the risk difference is also reasonable (large samples, $p$ and $q$ both not close to zero or one). Any of the three methods for computing the variance of $\hat{\theta}$ described above could be used in this approximation, so that we approximate $\hat{\theta} \sim N(\theta, \sigma^2)$ where $\sigma^2$ is treated as if fixed and known.

Those who might be reluctant  to use a normal approximation for $\hat{\theta}$ because it is also defined in terms of two estimated rate ratios in equation (\ref{eq2a}) (for which analysis is usually performed on the log scale) would probably be more willing to use normal approximations for sample proportions, and so be willing to approximate the distribution of $\hat{\theta}$ using ratios of two  normal distributions. Then, as Marsaglia (2006) points, out `approximations show that many of the ratios of normal variates encountered in practice can themselves be taken as normally distributed'. Hence normal approximations for $\hat{\theta}$ are not necessarily immediately unacceptable.

However, we have found by simulating $2 \times 2$ tables that the distribution of $\hat{\theta}$ often contains a skew tail. Hence normal approximations made directly for $\hat{\theta}$  are only very crude and  a better approximation for the sampling distribution of $\hat\theta$ is desirable. Here we develop such an approximation using a split lognormal distribution. This approximation will be used in our third model for random-effects meta-analysis (model 3) below.  Our split lognormal approximation simply uses  conventional normal approximations for the log risk ratios in (\ref{eq2a}) and then puts them together to approximate the distribution of $\hat{\theta}$.

Using the delta-method, we have approximately
\begin{equation}\ln(\hat{q}/\hat{p}) \sim N(\mu_1,\sigma_1^2)=N\left(\ln(q/p), \frac{1-q}{qN_2}+\frac{1-p}{pN_1}\right),\label{eq:a1}\end{equation}
using the notation in Table \ref{tab:1},
\ie $\ln(\hat{q}/\hat{p})$ is normally distributed with known mean and variance. Similarly, take $\ln ((1-\hat{q})/(1-\hat{p}))$ as
\begin{equation}\ln ((1-\hat{q})/(1-\hat{p})) \sim N(\mu_2,\sigma_2^2)=N\left(\ln((1-q)/(1-p)), \frac{q}{(1-q)N_2}+\frac{p}{(1-p)N_1}\right).\label{eq:a2}\end{equation}
These approximations work surprisingly well, and we use them both, so that we approximate
\begin{equation}
\label{eq:app}
X=\left\{\begin{array}{cl}
~\ln(1+\hat{\theta}) \sim N\left(\mu_1,\sigma_1^2\right) & \text{if $\hat{\theta} < 0$}\\
-\ln(1-\hat{\theta}) \sim N\left(-\mu_2,\sigma_2^2\right) & \text{if $\hat{\theta} \ge 0$}.
\end{array}
\right. \end{equation}
Our definition of $X$ in (\ref{eq:app}) ensures that $X$ is increasing in $\hat{\theta}$, so that our parameterisation makes computation as easy as possible.  
When using the approximation for $X$, and so $\hat{\theta}$, in (\ref{eq:app}), we follow the usual convention of taking the variances  $\sigma_1^2$ and $\sigma_2^2$ as known but we estimate them using $\hat{p}$ and $\hat{q}$ in practice. To use (\ref{eq:app}) in order to specify the approximate distribution of $\hat{\theta}$ in terms of ${\theta}$, all that is then required is to write $\mu_1$ and $\mu_2$ as functions of $\theta$. It is straightforward to write $\mu_1 = \ln(1+{\theta})$ when $\theta <0$, and $\mu_2 = \ln(1-{\theta})$ when ${\theta} \ge 0$. However it is not so straightforward to write $\mu_1$ and $\mu_2$ as functions of $\theta$ when $\theta  \ge 0$, and $\theta <0$, respectively, because then $\mu_1$ and $\mu_2$ are not directly specified by $\theta$. In the appendix we explain how to overcome this difficulty, so that the probability density function of $\hat{\theta}$ can be written in terms of $\theta$ when using the split normal approximation. We have found that our split lognormal approximation is able to capture the skew tail of the distribution of $\hat{\theta}$, and so is in general much more accurate than a  crude normal approximation.

We also explain how a variety of other inferences can be made for a single $2 \times 2$ table using the split normal approximation in the appendix, including the computation of confidence intervals and p-values. However for the purposes of using the split normal approximation in meta-analysis (model 3, below) we require only the probability density function of $\hat{\theta}$ as a function of $\theta$, so that likelihoods can be computed and the usual asymptotic theory of maximum likelihood applied.

\section{Random-effects meta-analysis}
We now present three different methods for performing random-effects meta-analyses using our measure. These three contrasting methods are all 2-stage methods because we require the computation of study specific data in the first stage that are then pooled in the second stage. For method 1 (the conventional random-effects model), in the first stage we compute the $\hat{\theta}_i$ and $\sigma_i^2$ (where the $\sigma_i^2$ may be computed using any of the above three approaches) and we pool these outcome data in stage two in the usual way. For method 2 we use the same outcome data (the $\hat{\theta}_i$ and $\sigma_i^2$) but avoid using a normal distribution, in order to avoid concerns about using this distribution for our measure that is constrained to lie within [-1,1]. Finally for method 3 we use the split-normal distribution to describe the within-study distributions and we include non-normal random-effects. When using method 3 in the first stage we compute the $\hat{\theta_i}$, and the variances $\sigma_{1i}^2$ and $\sigma_{2i}^2$, required in the study specific split lognormal within-study approximations from (\ref{eq:app}). In principle 1-stage meta-analyses (\eg Simmonds and Higgins 2006), that avoid within-study approximations and perform the analysis in a single stage, are possible and we return to this possibility in the discussion.

\subsection{Method one: The conventional random-effects method\label{meth1}}
This method is the simplest and most direct method: the study-specific outcome data, $\hat{\theta}_i$ and $\sigma^2_{i}$ are computed  and used directly as outcome data in the random-effects model for meta-analysis $\hat{\theta}_i \sim N({\theta}, \sigma^2_{i} + \tau^2)$. 
We refer to it as the `direct method' in the next section.
Any of the above three methods described above in section 2.3 for computing the $\sigma^2_{i}$ could be used in conjunction with this conventional approach. An advantage of this method is that, having computed the  $\hat{\theta}_i$ and $\sigma^2_{i}$, standard meta-analysis software packages can be used to perform the analysis. A variety of methods for estimating $\tau^2$ are available when using this standard approach (Veroniki {\em et al}, 2016). 

An advantage of using the new measure in this way is that there is no need to add $1/2$, or some other constant, to all counts to prevent infinities in $\hat{\theta}_i$ and $\sigma^2_{i}$. If both $p_i$ and $q_i$ are estimated as zero or unity, $\hat{\theta}_i$ would be zero as would its variance, and the study would be discarded; this is commonly done with 2-stage methods with conventional measures such as the odds ratio.

A potential problem here is that assuming a normal distribution for $\hat{\theta}_i$ is not especially appropriate, partly because $\hat{\theta} \in [-1,1]$, as discussed above. However standard random-effects meta-analyses are often performed using the risk difference, where this is also an issue but is not considered to be a sufficient concern to avoid this approach.
The next two methods address this problem. 
\subsection{Method two: a random-effects model using the beta distribution\label{meth2}}
The $\hat{\theta}_i$ lie in the interval [-1,1]. In order to use the beta distribution to model these estimates, we model the transformed outcome data $\hat{\psi}_i=(1+\hat{\theta}_i)/2$ so that $\hat{\psi}_i \in [0,1]$. Specifically, the observed $\hat{\psi}_i$ is taken as a random variable from the beta distribution with mean $\psi=(1+\theta)/2$ and variance $(\sigma_i^2+\tau^2)/4$, where  $\hat{\theta}_i$ and $\sigma_i^2$ are the same outcome data as in method one. Larger studies contribute more weight to the analysis via their smaller $\sigma_i^2$, as in the more conventional method above.

Thus the contribution to the likelihood from a study is
\[{\cal L}_i(\theta,\tau)=\hat{\psi}_i^{\alpha_i-1}(1-\hat{\psi}_i)^{\beta_i-1}/B(\alpha_i,\beta_i),\]
where $B$ denotes the beta function.  Here $\alpha_i$ and $\beta_i$ are the parameters of the beta distribution that models $\hat{\psi}_i$, and so we must parameterise $(\alpha_i,\beta_i)$ in terms of the mean $\psi=(1+\theta)/2$ and variance $(\sigma_i^2+\tau^2)/4$ corresponding to the $i$th study. Hence we take $(\alpha_i,\beta_i)$ to be the values that correctly provide these two moments of $\hat{\psi}_i$. This reparamerization of $(\alpha_i,\beta_i)$ to $\psi=(1+\theta)/2$ and $(\sigma_i^2+\tau^2)/4$ is easily performed and is described in the appendix.
The likelihood function is then the product of the study specific ${\cal L}_i(\theta, \tau)$ and approximate
inference is performed using the asymptotic theory of maximum likelihood.

Note that here we must add halves, or some other quantity, when there are zero counts or we
will obtain zero probabilities under the beta distribution.
The transformation $\hat{\psi}_i=(1+\hat{\theta}_i)/2$ is very simple so that inferences are easily back-transformed to the original scale.

\subsection{Method three: a random-effects method using the split-lognormal approximation\label{meth3}}
Here the split lognormal approximation, from  equations (\ref{eq:a1}),  (\ref{eq:a2}) and  (\ref{eq:app}) is used to model the within-study distributions. As explained in section 2.4, we can specify the approximation in terms of the GRRR. We can therefore apply this approximation to each study, and in terms of its study specific true underlying effect $\theta_i$. If we assume a common-effect model ($\tau^2=0) $, so that $\theta_i=\theta$ for all $i$, inference is easily formed using maximum likelihood as for the previous model where the likelihood function is the product of the ${\cal L}_i({\theta_i}) = {\cal L}_i({\theta})$. Larger studies contribute more weight to the analysis via their smaller $\sigma_{1i}^2$ and  $\sigma_{2i}^2$ used in their study specific split lognormal approximation (equations \ref{eq:a1},  \ref{eq:a2} and  \ref{eq:app}).

To include a random effect, and so fit a random-effects model, we  take $\psi_i=(1+\theta_i)/2$ to have a beta distribution, with mean $\psi=(1+\theta)/2$ and variance $\tau^2/4$. Written in terms of $\psi_i$, the within-study likelihood ${\cal L}_i({\theta_i})$ is ${\cal L}_i({2\psi_i}+1)$. This within-study likelihood is then integrated over the distribution of $\psi_i$ in order to integrate out the random-effects in the usual way, so that
\[{\cal L}(\theta,\tau)=\frac{\int_0^1 {\cal L}_i(2\psi_i-1)\psi_i^{\alpha-1}(1-\psi_i)^{\beta-1}\dee \psi_i}{B(\alpha,\beta)}.\]
where this integration is performed numerically. As in method 2, approximate
inference is performed using the asymptotic theory of maximum likelihood. Here $\alpha$ and $\beta$ are chosen so that $\text{E}(\psi_i)=\psi=(1+\theta)/2$ and $\text{var}(\psi_i)=\tau^2/4$. 
This reparameterisation is similar to the one used in method 2 and is also described in the appendix. 

This is a similar approach to modelling the random effect as in the previous method. A conceptual difference is that in method 2 we used a beta distribution to model the $\hat{\psi}_i$ but 
here we instead use this distribution to model the true underlying $\psi_i$; we assume that the $\psi_i$ follow a common distribution so that the same $\alpha$ and $\beta$ are used for all studies when computing the likelihood. 
In this sense method 3 is computationally simpler, but it requires a separate numerical integration for every study when computing the likelihood. Hence method 3 is the most computationally expensive of the three methods that we propose,
but uses the most realistic model.

\section{Application to meta-analysis datasets}

In this section we will use 3 real examples to illustrate the use of our measure in practice. The first example involves thirteen randomized control trials from 1948 to 1976 on the prevention of tuberculosis using the BCG (Bacillus Calmette-Gu\'{e}rin) vaccine, with data given in Hartung, Knapp and Sinha (2008), but taken originally from Colditz {\em et al} (1994).  The event of interest is contracting tuberculosis. The second example involves 22 trials of streptokinase following mycocardial infarction, given in Egger, Altman and Smith (2001). 
Briefly, from 1959, 21 trials were carried out
to see whether streptokinase could reduce 6-month mortality from infarction; this was feasible because streptokinase can dissolve blood clots.  Here the event of interest is death. The third example one of the eleven randomised control trials of lamotrigine  (Ramaratnam, Panabianco and Marson, 2016), from 1989 to 2007, as an adjunctive therapy for the treatment of drug-resistant partial epilepsy. The outcome is a 50\% or more reduction in seizure frequency.

Table \ref{tab:40} shows the results from using all three methods described in section 3 on our main three examples using maximum likelihood estimation and the asymptotic theory of maximum likelihood to make inferences.
The within-study variance was computed exactly (by enumeration) for methods 1 and 2.
Table \ref{tab:dsl} shows the results using the two stage method described in section \ref{meth1},
instead using the the Dersimonian and Laird (1986) method. In Table \ref{tab:dsl} the quoted $\hat{\tau}$ is the square root of the corresponding DerSimonian and Laird estimate $\hat{\tau}^2$. It can be seen that the results using the more conventional DerSimonian and Laird method are very similar to those using our `direct' method.
\begin{table}[h]
\begin{tabular}{|l|c|l|l|l|l|l|} \hline
Analysis method&Dataset & $\hat\theta$ & s.e. & $\hat{\tau}$ & s.e.  \\ \hline
2-stage Direct (sec \ref{meth1})&TB &-.496&.088&.292&.066\\ \hline
2-stage Beta (sec \ref{meth2})&TB & -.489&.083&.270&.061\\ \hline
2-stage Lognormal (sec \ref{meth3})&TB & -.505&.075&.239&.053\\ \hline
2-stage Direct (sec \ref{meth1})&Strept&-.170&.045&.149&.043\\ \hline
2-stage Beta (sec \ref{meth2})&Strept&-.174&.045&.152&.041\\ \hline
2-stage Lognormal (sec \ref{meth3})&Strept&-.200&.023&0&0\\ \hline
2-stage Direct (sec \ref{meth1})&Lamot&~.201&.036&.078&.031\\ \hline
2-stage Beta (sec \ref{meth2})&Lamot&~.200&.036&.031&.060\\ \hline
2-stage Lognormal (sec \ref{meth3})&Lamot&~.132&.030&.049&.046 \\ \hline
\end{tabular}
\caption{\label{tab:40}Results for the new measure applied to three examples, using all three methods.}
\end{table}

\begin{table}[h]
\begin{tabular}{|l|c|l|l|l|l|l|} \hline
Analysis method &Dataset & $\hat\theta$ & s.e. & $\hat{\tau}$ & $I^2$\% \\ \hline
2-stage Direct (sec \ref{meth1})&TB &-.493 & .102&.345&97.6\\ \hline
%2-stage Lognormal (sec \ref{meth2})&TB &-.533&.110&.368&97.0\\ \hline
2-stage Direct (sec \ref{meth1})&Strept&-.168&.041&.133&63.0\\ \hline
%2-stage Lognormal (sec \ref{meth2})&Strept&-.156&.041&.124&57.5\\ \hline
2-stage Direct (sec \ref{meth1})&Lamot&.202&.033&0&0\\ \hline
%2-stage Transform (sec \ref{meth2})&Lamot&.194&.033&0&0 \\ \hline
\end{tabular}
\caption{\label{tab:dsl}Results for the new measure applied to parallel studies, using the Dersimonian and Laird method
and assuming a normal distribution for $\hat\theta$.}
\end{table}

The outcomes are  harmful (death and contracting tuberculosis)  in our first two examples and beneficial (reduction in seizure frequency) in our third example. Hence $\theta<0$ indicates treatment benefit in the first two examples and $\theta>0$ indicates treatment benefit in the third example. Using the results in Table \ref{tab:40}, and normal approximations for the maximum likelihood estimates, we infer that the treatment is beneficial in all three examples. Our use of the GRRR then allows us to communicate these findings to a lay audience in a very simple and direct way. For example, let us take $\hat{\theta}=-0.5$  from Table \ref{tab:40} for our first example (TB). We are then able to tell a lay audience that one way to explain the extent of the treatment efficacy is to say that we estimate that 50\% (i.e. half) of the population who do not take the vaccine and contract TB would also contract TB if they instead took the vaccine (where we assume that all those who would not contract TB without taking the vaccine also would not contract this if they
took the vaccine). In other words, we estimate that around half the population who do not take the vaccine and contract TB would instead avoid contracting this if they had taken the vaccine. Statements such as these nicely convey the notion that the vaccine has real benefit (but is not perfect) in a simple way, whilst being statistically principled. We can also quantify the uncertainty in this statement.  From a 95\% confidence interval for $\theta$ this, the 50\% that we quoted could in fact be between around 30\%-70\%.

As another illustration, let us take $\hat{\theta}=0.2$ from Table \ref{tab:40} for our third example. We are then able to tell a lay audience that one way to explain the extent of the treatment efficacy is to say that, in addition to those who experience notable seizure frequency in the reduction without the lamotrigine (and would also experience this if they took this treatment), a further 20\% of those who do not experience notable seizure frequency without lamotrigine would instead experience this if they took the treatment. However there is uncertainty in this estimate and (from a 95\% confidence interval for $\theta$) this percentage could be between around 10\%-30\%. Again, these statements clearly convey the potential benefit of taking the treatment in an especially direct and transparent manner. 

A final point is that analysts may be reluctant to use our proposed measure in analysis because it is unconventional, but may wish to convert results using other measures to it, so that explanations such as these can be given. This conversion can be performed upon adopting a representative baseline risk $p$ for the control group and we give full details of the calculation required in the appendix for converting the odds ratio in this way. We return to this issue in the discussion.

\section{Discussion}
A new treatment effect measure (generalized relative risk reduction, or `GRRR'), based on relative risk, has been introduced.
The new measure gives a treatment effect on a scale from minus one to plus one, with zero indicating no treatment effect.
There is a clear causal interpretation that accompanies the new measure and that can be used to communicate the results to those with little or no formal statistical training. Those who are faced with explaining the findings from meta-analyses, and statistical analyses more generally, to the general public are likely to find our new measure especially useful.
This may include journalists and politicians as well as clinicians. Health economists may also find this measure useful,
as costs of using or not using a new intervention are straightforward to calculate. We suggest that interpretations using our measure could be included in `plain language summaries' that accompany Cochrane reviews and other information sources that are intended for a wide audience. For the `take-home' messages from statistical analyses to be fully appreciated by the general public we require methods such as those that we present here. We hope that, at the very least, our methods will provide further ideas for communicating statistical findings in a simple and direct, and yet still statistically principled, manner that is widely accessible.   

We have developed three methods for performing 2-stage random-effects meta-analysis that use our new measure.
We therefore have proof of concept that it may be used in conjunction with quite sophisticated statistical models. Future work could focus on other types of models where binary outcome data are modelled, such as logistic regressions and generalised linear mixed models. For example, regression modelling could be carried out by allowing $\psi =(1+\theta)/2\in [0,1]$ to be
a logistic function of  covariates. These possibilities for more complex modelling include 1-stage methods for random-effects meta-analysis and the authors have developed two further methods of this type. These 1-stage methods have been found to produce similar results to the 2-stage methods presented here and may form the subject of future work. We have also performed a small scale empirical investigation to determine if the proposed measure results in better model fits than the more conventional (log) odds ratio. Further investigation is needed but our preliminary investigation suggests that models using our measure describe real meta-analysis datasets just as well as more conventional measures of treatment effect. 

Rather than motivate our new measure as providing better fitting models to data, we have proposed it primarily so that the resulting statistical inferences can be more easily communicated to general audiences.  We suggest therefore that it is particularly suited to providing results that could be communicated in plain language summaries such as those used by Cochrane. It is most straightforward to report results using our measure after actually using our measure as the outcome in analysis, but we suspect that many analysts would object to this idea unless our measure becomes more widely used and accepted. We would encourage analysts who might be uncomfortable in using our new measure in analysis to consider converting their results using a measure of their choice to ours, so that conclusions using our measure can  be reported despite the fact that an alternative measure was used in analysis. 

A classical (frequentist) approach to statistical inference has been adopted here but the likelihood-based methods can also be used for Bayesian inference.
Prior distributions for all parameters would then be needed.
The main difficulty is determining a suitable prior distribution for $\theta$,
and the beta distribution for $\psi$ is an obvious candidate. It can be used when there is a lot of prior information, and also includes as special cases the uniform and Jeffreys priors.
Markov chain Monte Carlo (MCMC) would probably be used to perform analyses because the resulting posterior distributions are unlikely to be analytically tractable.

The only other candidate measure of treatment effect for $2 \times 2$ tables that is so easily interpretable is the number needed to treat. However this measure has unacceptable statistical properties. We suggest that the GRRR is suitable as a replacement for this measure as it is both easily interpretable and
has acceptable statistical properties. The GRRR has however two undesirable properties: it is not invariant when the treatment and control groups are interchanged, as we have explained, and furthermore $\theta$ is not differentiable when $\theta=0$. This latter property has not caused us any problems here but this could result in difficulties for the unwary. At the very least, our proposed measure has much better properties than the number needed to treat, which we regard as its main contender for an easily interpreted and intuitively appealing measure of treatment effect for comparative binary outcome data.

To summarise, we hope that meta-analysts, and indeed the the  statistics community more generally, will be convinced by the case for our new measure of treatment effect, and that they will find it to be a useful new way to measure and communicate the results from comparative trials that involve a binary event of interest. We also hope that our work will serve to stimulate debate about the best way to communicate statistical conclusions to those with little or no formal statistical training.

\section{Appendix: detailed formulae}
\subsection{The split-lognormal distribution}
In section 2.4 of the main paper we explain that it is possible to write the probability density function (pdf) of $\hat{\theta}$ in terms of $\theta$. In this section of the appendix we give full details about how this is done.

To derive an approximation for the sampling distribution of $\hat\theta$, using the delta method and assuming normality of the logged risk ratio, we take $\ln(\hat{q}/\hat{p})$ as
obeying (8) and $(1-\hat{q})/(1-\hat{p})$  as obeying (9), where these equations are in the main paper.
Then $\hat{q}/\hat{p}$ follows a lognormal distribution. Similarly, $(1-\hat{q})/(1-\hat{p})$ follows a lognormal distribution, and $\hat\theta$ is as defined in (10) via $X$.
From this approximation for $X$, and hence $\hat{\theta}$, the  pdf and distribution function of $\hat{\theta}$ can be obtained. Hence
p-values, confidence  intervals, and the moments of this distribution may also be computed.
\subsection{The probability density function and the cumulative distribution function of $\hat{\theta}$}
We can identify four cases from equation (10) and we evaluate the pdf for each case. Then the pdf is then defined for all cases. These four cases arise because  $\hat{\theta}$ can take either sign (we use normal approximations and so the probability that $\hat{\theta}=0$ is zero), and $\theta$ can be of the same sign as
$\hat{\theta}$ or not. To see why we consider these four cases separately, consider the first line in the right hand side of (10), which applies to $\hat{\theta}<0$. If $\theta<0$, so that  $\hat{\theta}$ and $\theta$ are both negative, then it is straightforward to write $\mu_1=\ln(1+\theta)$ and a normal approximation for $X$, and hence $\hat{\theta}$, is immediate for the combination of $\hat{\theta}<0$ and $\theta<0$. However if $\hat{\theta}<0$ and  $\theta \ge 0$ then we no longer have  $\mu_1=\ln(1+\theta)$ and the parameterisation of the normal approximation is not quite so immediate. Following a similar argument for the second line in the right hand side of (10), we can see that the difficulties occur when $\hat{\theta}$ and $\theta$ are not of the same sign, and so we adopt a `divide and conquer' approach of considering each case separately.

When $\hat{\theta} < 0$, $\theta < 0$, we have the
approximation $\ln(1+\hat{\theta}) \sim N[\mu_1=\ln(q/p)=\ln(1+\theta),\sigma_1^2]$, and so
the pdf
\[f(\hat{\theta})=\frac{\exp\{-(\ln(1+\hat{\theta})-\ln(1+\theta))^2/2\sigma_1^2\}}{\sqrt{2\pi\sigma_1^2}(1+\hat{\theta})}.\]
The corresponding distribution function is
\begin{equation}\Phi\{(\ln(1+\hat{\theta})-\ln(1+\theta))/\sigma_1\},\label{eq:phi1}\end{equation}
where $\Phi(\cdot)$ is the standard normal distribution function.
Here we have used the fact that $\mu_1=\ln(q/p)=\ln(1+\theta)$. However, when $\theta \ge 0$, $\mu_1$ is not specified by $\theta$.
This is because $\theta=1-(1-q)/(1-p)$ when $\theta \ge 0$ from which $q/p$ cannot be determined.
However, a substitute value can be found by requiring that the total probability is unity.
We therefore use the fact that
\[\text{Prob}(\hat{q}/\hat{p} > 1)=\Phi(\mu_1/\sigma_1)=\text{Prob}((1-\hat{q})/(1-\hat{p})) < 1=\Phi(-\mu_2/\sigma_2)\]
to obtain the useful result
\begin{equation}\mu_1/\sigma_1=-\mu_2/\sigma_2\label{eq:trick}\end{equation}
from which, because $\mu_1=\ln(q/p)$, the approximate mean of $\ln(1+\hat{\theta})$ for $\hat{\theta} < 0$ can be found
from $\mu_2=\ln((1-q)/(1-p))$, which is also equal to $\ln(1-\theta)$ when $\theta \ge 0$.
Using (\ref{eq:trick}), when $\hat{\theta} < 0$, $\theta \ge 0$
\[f(\hat{\theta})=\frac{\exp\{-(\ln(1+\hat{\theta})+(\sigma_1/\sigma_2)\ln(1-\theta))^2/2\sigma_1^2\}}{\sqrt{2\pi\sigma_1^2}(1+\hat{\theta})}.\]
The corresponding distribution function is
\[\Phi\{\ln(1+\hat{\theta})/\sigma_1+\ln(1-\theta)/\sigma_2\}.\]
Similarly, when $\hat{\theta} \ge 0$, we have for $\theta \ge 0$ that
\[f(\hat{\theta})=\frac{\exp\{-(\ln(1-\hat{\theta})-\ln(1-\theta))^2/2\sigma_2^2\}}{\sqrt{2\pi\sigma_2^2}(1-\hat{\theta})}.\]
The corresponding distribution function is
\begin{equation}\Phi\{(-\ln(1-\hat{\theta})+\ln(1-\theta))/\sigma_2\},\label{eq:phi3}\end{equation}
where the minus sign arises because $\ln(1-\hat{\theta})$ is a decreasing function of $\hat\theta$.
Finally for $\hat{\theta} \ge 0 $, $\theta < 0$
\[f(\hat{\theta})=\frac{\exp\{-(\ln(1-\hat{\theta})+(\sigma_2/\sigma_1)\ln(1+\theta))^2/2\sigma_2^2\}}{\sqrt{2\pi\sigma_2^2}(1-\hat{\theta})}.\]
The corresponding distribution function is
\[\Phi\{-\ln(1-\hat{\theta})/\sigma_2-\ln(1+\theta)/\sigma_1\}.\label{eq:phi4}\]

\subsubsection{P-values and confidence intervals}
At the end of section 2.4 of the main paper we explain that other inferences can be made for a single $2 \times 2$ table. In this section of the appendix we give full details of this.

First, p-values are given when $\theta=0$ so that $\mu_1=\mu_2=0$. The 1-sided p-value for obtaining $\hat\theta$ at least as large as observed when $\hat{\theta} \ge 0$ is $\Phi(\ln(1-\hat{\theta})/\sigma_2)$,
using (\ref{eq:phi3}) and the identity $1-\Phi(x)=\Phi(-x)$.
The 1-sided p-value for obtaining $\hat\theta$ at least as negative as observed when $\hat{\theta} < 0$ is $\Phi(\ln(1+\hat{\theta})/\sigma_1)$ from (\ref{eq:phi1}).
From (\ref{eq:phi1}) and (\ref{eq:phi3}) we obtain the formula for the corresponding 2-sided p-values, \ie for $\hat\theta$ to exceed the observed $|\hat{\theta}|$ in either direction.
This is
\[p=\Phi(\ln(1-|\hat{\theta}|)/\sigma_1)+\Phi(\ln(1-|\hat{\theta}|)/\sigma_2)\]
from which we can see that the 2-sided p-value is 1 if $\hat{\theta}$ is exactly zero (this is impossible using normal approximations but could arise in real data.)

Confidence intervals for $\theta$ can  be computed by equating quantiles of the pdf to the required values.
When $\hat{\theta} \ge 0$, from (\ref{eq:phi3}) and the corresponding pdf we obtain the size $\alpha$ limits for $\ln(1-\theta)$ as $\ln(1-\hat{\theta}) \pm \sigma_2 z_{\alpha/2}$,
where $z_\alpha$ is the $100\alpha$ percentile of the normal distribution. From this the limits for $\theta$ are
$\theta=1-(1-\hat{\theta})\exp(\pm \sigma_2 z_{\alpha/2}$).
However this calculation assumes that
$\hat{\theta} \ge 0$ and $\theta \ge 0$ but it may happen that lower limit is negative so that we have $\hat{\theta} \ge 0$ and $\theta < 0$
at the lower end of the confidence interval. In this case, the lower limit must be recomputed as
\[\theta=(1-\hat{\theta})^{-\sigma_1/\sigma_2}\exp(-\sigma_1 z_{\alpha/2} )-1.\]
This follows by using (\ref{eq:trick}) to deal with the issue that the signs of $\hat{\theta}$ and ${\theta}$ are not the same and proceeding in a similar way as when deriving the pdf and cumulative distribution function.

Similarly, when $\hat{\theta} < 0$, the confidence interval is $\theta=(1+\hat{\theta})\exp(\pm \sigma_1 z_{\alpha/2})-1$, unless the upper limit is positive, in which case it should be recomputed as
$\theta=1-(1+\hat{\theta})^{-\sigma_2/\sigma_1}\exp(\sigma_2 z_{\alpha/2})$.

\subsection{The variance of $\hat{\theta}$}
In order to use standard methods for meta-analysis we require within-study variances.  In section 2.3.3. of the main paper we discuss an approximate formula for this. In this section of the appendix we derive this formula.

A large-sample approximation for the variance of $\hat\theta$ has been developed, and works very well when $0.1 < p < 0.9$, $0.1 < q < 0.9$, and $N_1 > 100, N_2 > 100$.
Because of the tractability of the lognormal distribution, one can calculate $A_n=\text{E}_L\{(\hat{q}/\hat{p})^n\}$, where the integral for the expectation
is truncated at $\hat{q}/\hat{p}=1$, so that the expectation is calculated over the range where $\hat{q} < \hat{p}$, \ie
\[A_n=\frac{1}{\sqrt{2\pi \sigma_1^2}}\int_{0}^1 x^{n-1}\exp(-(\ln(x)-\mu_1)^2/2\sigma_1^2)\dee x.\]
This integral can be evaluated analytically by changing variable to $y=\ln(x)$, so that the integration is now performed over $(-\infty, 0)$, and completing the square in the exponent.
Similarly $B_n=\text{E}_R\{((1-\hat{q})/(1-\hat{p}))^n\}$ can be calculated as
\[B_n=\frac{1}{\sqrt{2\pi \sigma_2^2}}\int_{0}^1 x^{n-1}\exp(-(\ln(x)-\mu_2)^2/2\sigma_2^2)\dee x\]
where the integral for the expectation is truncated at $(1-\hat{q})/(1-\hat{p}) =1$, so that the expectation is calculated over the range where  $1-\hat{q} \le 1-\hat{p}$, or $\hat{q} \ge \hat{p}$. Evaluating the integrals analytically gives
\[A_n=(q/p)^n\exp(n^2 \sigma_1^2/2)\Phi(-\mu_1/\sigma_1-n\sigma_1),\]
\[B_n=((1-q)/(1-p))^n\exp(n^2 \sigma_2^2/2)\Phi(-\mu_2/\sigma_2-n\sigma_2).\]
Hence we have one approximation for $\hat{q} < \hat{p}$ (the `left' side of the distribution), and a different approximation for $\hat{q} \ge \hat{p}$ (the right side).
From the definition of $\hat{\theta}$ we then have
\[\text{E}(\hat{\theta})=\text{E}_L(\hat{q}/\hat{p}-1)+\text{E}_R(1-(1-\hat{q})/(1-\hat{p}))=A_1-A_0+B_0-B_1,\]
\[\text{E}(\hat{\theta}^2)=A_2-2A_1+A_0+B_2-2B_1+B_0,\]
from which the variance of $\hat\theta$ can be computed as $\text{E}(\hat{\theta}^2) - \text{E}(\hat{\theta})^2 $, on replacing $p, q$ in the formula by $\hat{p}, \hat{q}$ respectively. Note that we have evaluated the expectation of $\hat{\theta}$ from its definition, where we have evaluated this expectation by integrating over the two areas of the sample space separately.
This approximation to $\sigma^2$ is surprisingly accurate.
It requires the computation of the normal distribution function 6 times.
\section{Reparameterising the beta distribution}
As explained in sections 3.2 and 3.3, to use methods 2 and 3 we need to specify the mean $\psi$ and variance $\sigma^2$ of a beta distribution, and then compute the usual beta function parameters $\alpha, \beta$
within the section of computer code that computes the log-likelihood function. We give the details here. For method 2 this mean is $\psi=(1+\theta)/2$ and the variance is $\sigma^2= (\sigma_i^2+\tau^2)/4$; for method 3 the variance is instead $\sigma^2=\tau^2/4$.

 We have the standard result
\[\psi=\alpha/(\alpha+\beta),\]
\[\sigma^2=\frac{\alpha\beta}{(\alpha+\beta)^2(\alpha+\beta+1)}.\]
Hence $\sigma^2=\frac{\psi(1-\psi)}{\alpha+\beta+1}$, so that
\[\alpha=\psi\{\frac{\psi(1-\psi)}{\sigma^2}-1\},\label{eq:alpha}\]
\[\beta=(1-\psi)\{\frac{\psi(1-\psi)}{\sigma^2}-1\}.\label{eq:beta}\]

We must constrain our model parameters so that $\alpha$ and $\beta$ are positive in this reparameterisation. This could be a problem  if the function minimiser chooses very large values of $\sigma^2$ and further reparameterisations could be used to make the numerical methods more robust.

\subsection{Converting to our new measure}

As explained at the end of section 4, we anticipate that some applied analysts may not be convinced by the  case for using our measure in statistical analysis, but despite this will find the new measure to be an attractive option for communicating their findings to those with little or no formal statistical training. To use our proposed measure to communicate findings when alternative measures of treatment effect have been used in  analysis, we need ways to convert other measures to ours. In this section we explain how to convert the odds ratio to the GRRR. The methods and issues are similar when converting other measures of treatment effect.

When converting from one measure of treatment effect to another, for example the odds ratio to the risk difference, we need to take a representative baseline risk $p$. This can be the average value of $p$ for the studies in the meta-analysis, either unweighted or weighted. We can then use the implied $q$ from one measure of treatment effect to compute the other measure that we wish to convert to. By taking into account the uncertainty in the first of these measures, we can communicate the uncertainty in the conversion.

Let us start by converting the odds ratio to a finite range. This can be done in several ways,
the simplest arguably being
\begin{equation}\phi=\frac{OR-1}{OR+1}=\frac{q-p}{p+q-2pq}.\label{eq:ortheta}\end{equation}
The measure in (\ref{eq:ortheta}), which itself is easily derived from the definition of the odds ratio, can then be converted into $\theta$ defined by equation (2) of the main paper. On eliminating $q$,
\begin{equation}\theta=\left\{\begin{array}{ll}
\frac{2(1-p)\phi}{1-\phi+2p\phi} & \text{if $\phi < 0$}\\
\frac{2p\phi}{1-\phi+2p\phi} & \text{if $\phi \ge 0$}.
\end{array}
\right. \label{eq:back}\end{equation}
so that the OR can easily be converted to $\phi$, which can then be converted to our measure $\theta$.

\end{document}